\documentclass[prl,aps,showpacs,twocolumn,preprintnumbers,
amsmath,amssymb,superscriptaddress,longbibliography]{revtex4-1}
\usepackage[english]{babel}
\usepackage{amsmath,amssymb,amsfonts}
\usepackage{graphicx}
\usepackage[colorlinks=True,linkcolor=red,citecolor=blue,urlcolor=blue]{hyperref}
\usepackage{bookmark}
\usepackage{xcolor}
\usepackage{chngcntr}
\usepackage{braket}
\usepackage{nicefrac}
\usepackage{bm}
\usepackage{bbm}
\usepackage{braket}
\usepackage{slashed}
\usepackage[normalem]{ulem}
\usepackage{array}
\newcolumntype{C}{>{$}c<{$}}

\renewcommand{\dag}{^{\dagger}}

\newcommand{\diff}{\mathrm{d}}

\newcommand{\abs}[1]{\left\lvert #1 \right\rvert}

\allowdisplaybreaks

\renewcommand{\dag}{^{\dagger}}

\usepackage[sectionbib]{bibunits}
\begin{document}
\defaultbibliographystyle{apsrev4-1} 
\defaultbibliography{Biblio} 

\title{
Anomalous {dynamical} Casimir effect in an expanding ring
}

\author{Baptiste Bermond}
\email{baptiste.bermond@ens-lyon.fr}
\affiliation{ENSL, CNRS, Laboratoire de Physique, F-69342 Lyon, France}

\author{Adolfo G. Grushin}
\email{adolfo.grushin@neel.cnrs.fr}
\affiliation{Univ. Grenoble Alpes, CNRS, Grenoble INP, Institut N\'eel, 38000 Grenoble, France}

\author{David Carpentier}
\email{david.carpentier@ens-lyon.fr}
\affiliation{ENSL, CNRS, Laboratoire de Physique, F-69342 Lyon, France}

\begin{abstract}
The Casimir effect is a macroscopic evidence of the quantum nature of the vacuum. On a ring, it leads to a finite size correction to the vacuum energy. In this work, we show that this vacuum’s energy and pressure acquire additional, sizable corrections, when the ring's radius is increased fast enough, an experimentally accessible model of an expanding universe.  This effect is distinct from the dynamical Casimir effect: it is a manifestation of the conformal anomaly, originating from the spacetime curvature induced by the increase of the ring’s radius. This anomalous dynamical Casimir effect is measurable through the work necessary to increase the ring size, which becomes non-monotonous in time.
\end{abstract}
\date{\today}
\maketitle 
The Casimir effect is one of the most striking macroscopic manifestations of the quantum nature of vacuum~\cite{casimir1948attraction}.
It emerges when this vacuum is subject to boundary conditions, 
\textit{e.g.} two metallic plates separated by
an electromagnetic vacuum experience an attractive Casimir force~\cite{bordag2009advances,mostepanenko1997casimir,bressi2002measurement}. 
This effect has been tested experimentally using 
metals in vacuum\cite{bressi2002measurement,PhysRevLett.78.5,PhysRevLett.81.4549} and turned to be repulsive by replacing vacuum by a fluid~\cite{Munday2009}.
The Casimir effect generalizes to the
case of dielectric plates \cite{lifshitz1992theory,Klimchitskaya2009,Woods2016}, liquid crystals \cite{ajdari1992pseudo}, 
metamaterials\cite{Boyer1974,Kenneth02,Henkel_2005,Leonhardt_2007,Rosa08}, liquids and gases \cite{PhysRevB.98.201408}, and 
relativistic \cite{zhabinskaya2008casimir,shytov2009long,lu2021casimir}
and topological~\cite{Grushin2011a,Grushin2011b,Rodriguez-Lopez2014} matter.
When the boundary conditions are time-dependent, 
the increase of vacuum energy and associated properties lead to the so-called dynamical Casimir effects~\cite{moore1970quantum,dodonov2020fifty,wilson2011observation,jaskula2012acoustic,lahteenmaki2013dynamical,johansson2009dynamical,vezzoli2019optical}. 

The Casimir effect also affects groundstate properties in a finite geometry without boundaries. For example, when 
 a massless field is confined in a ring of size $L$ in dimension $D=1$, the finite-size energy correction 
is a manifestation of the Casimir effect \cite{francesco2012conformal}. 
In this paper, we consider the fate of this Casimir effect when the size of this 
ring is rapidly increased in time. 
Such a system is an analog of a 1D universe in expansion. 
Analogous expanding universes can be implemented experimentally for example 
at the edge of a quantum Hall sample with a time-dependent geometry~\cite{nambu2023analog}, 
and in Bose-Einstein condensate by modulating either their 
shape~\cite{fedichev2004cosmological,weinfurtner2005analogue,eckel2018rapidly,llorente2019expanding} or their scattering length~\cite{jain2007analog}. 
\begin{figure}
    \centering
    \includegraphics[width=0.8\columnwidth]{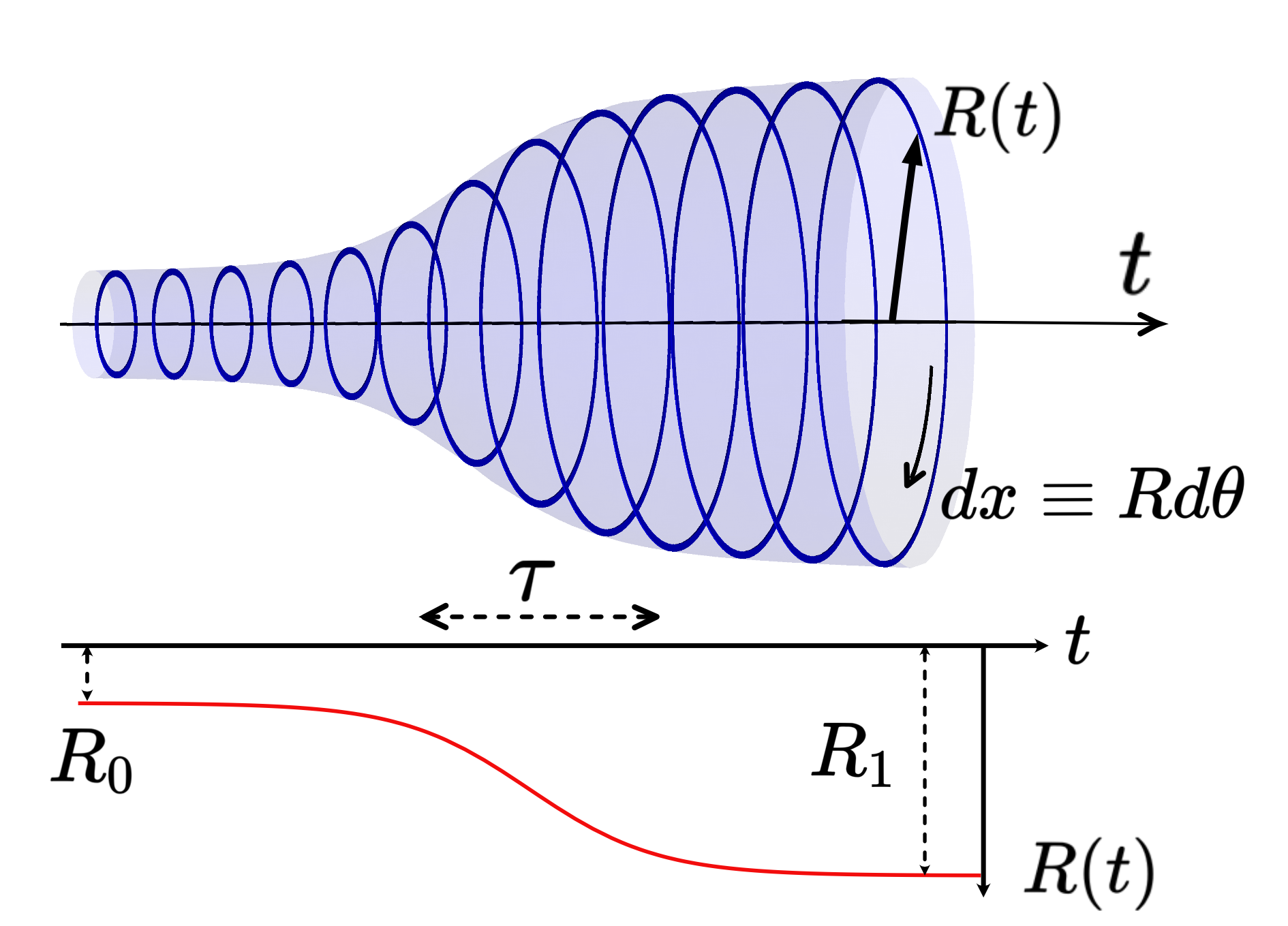}
    \caption{{Analog expanding universe along a ring of radius $R(t)$.}
    This radius is driven from $R_0$ to $R_1$ over a time $\tau$.}
    \label{fig:expanse}
\end{figure}
In this work we show that when the expansion of the ring is fast enough the Casimir effect has an additional contribution, distinct from the dynamical Casimir effect resulting from dynamical boundary conditions. 
{Because of this difference, and its link to quantum anomalies~\cite{bertlmann2000anomalies} that we now introduce, we refer to it as the anomalous dynamical Casimir effect}. 
The onset scale of this effect is set by the curvature of spacetime which modifies the ground state properties. 
In the present case, this curvature of spacetime originates from the time dependence of the spatial curvature of the ring, {\it i.e.} its radius. 
In practice, we account for the modifications of the vacuum properties by this spacetime curvature 
through the conformal anomaly of the quantum field theory for the energy density and pressure, 
and the gravitational anomaly for the energy current and momentum \cite{bertlmann2000anomalies}. 
We stress that the curved spacetime description of the  dynamical Casimir 
 effect~\cite{moore1970quantum,fulling1976radiation} relies on a
 metric with a vanishing curvature~\cite{davies1977quantum}, unrelated to a conformal anomaly. 
Hence, the anomalous {dynamical} Casimir effect, originating from quantum field theory anomalies, is thus distinct from the dynamical Casimir effect in a quantum wire with a varying length or boundary conditions.

In contrast, the anomalous dynamical Casimir effect has a similar origin to that of the Hawking or Unruh radiations. 
All of these phenomena originate from 
a curvature of spacetime~\cite{christensen1977trace,robinson2005relationship,banerjee2008hawking,banerjee2009hawking,gim2015quantal,bermond2022anomalous}, 
originating from either a spatial variation of the metric~\cite{unruh1981experimental} for the Hawking effect, or a dynamical origin for both the Unruh effect and the present anomalous dynamical Casimir effect. 
The anomalous dynamical Casimir effect opens an alternative to test the effect of a spacetime curvature on the vacuum properties, in addition to the experimental realization of the Hawking effect ~\cite{unruh1981experimental,garay2000sonic,garay2001sonic,unruh2007quantum,barcelo2001analogue,jain2007quantum,macher2009black,steinhauer2014observation,steinhauer2016observation}.
While the experimental signatures of these phenomena typically rely on the measurement of the particle density and its correlations~\cite{jain2007quantum,steinhauer2014observation,steinhauer2016observation,johansson2009dynamical,jaskula2012acoustic}, 
here we show that a remarkable feature of the anomalous dynamical Casimir effect appears as a mechanical instability, related to the behavior of the system's radiative pressure $p$.


We consider a massless field of excitations moving at a velocity $c_s$ along a ring of radius $R(t)$.  
Accounting for the expansion of the radius $R(t)$ (or a variable velocity $c_s$ as detailed in the supplemental material (SM)~\cite{SuppMat}),  
we describe such an expanding ring using a Friedmann–Lema{\^i}tre–Robertson–Walker metric \cite{LACHIEZEREY1995135,davies1977quantum}: 
\begin{equation}
    \diff s^2=c_s^2\diff t^2-R^2(t)\diff \theta^2 = c_s^2\diff t^2-a^2(t)\diff x^2\,,
\label{eq:metric}
\end{equation}
where $\diff x=R_0\diff \theta$, and the scale factor $a(t)=R(t)/R_0$, see Fig.~\ref{fig:expanse}. 
Analogs of such an expanding universe have been proposed when considering the azimuthal mode of a ring of a Bose-Einstein
condensate submitted to a forced radial expansion \cite{weinfurtner2005analogue,jain2007analog,eckel2018rapidly,llorente2019expanding}.

In a static finite system with $R(t)=R_0$ and in the low temperature regime such that 
$k_BT\ll \hbar c_s/R$, the thermodynamic properties of a massless relativistic system are controlled by 
a single energy scale,  the Casimir energy density $ \varepsilon_{\mathcal{C}}$, set by the size $L=2\pi R$. 
In particular the energy density and pressure satisfy \cite{francesco2012conformal}
\begin{equation}
    \varepsilon=p= - \frac{\mathcal{C}_w}{2} \varepsilon_{\mathcal{C}}\, ; \, 
     \varepsilon_{\mathcal{C}}= \frac{\pi \hbar c_s}{6L^2} = \frac{\hbar c_s}{24\pi  R^2}  \, ,
\label{eq:CasimirEnergy}
\end{equation}
with $\mathcal{C}_w=c_+ + c_-$ where the $c_\pm$ are the central charges of the left and right moving excitations of the theory.

\begin{figure*}[th]
    \centerline{\includegraphics[width=\textwidth]{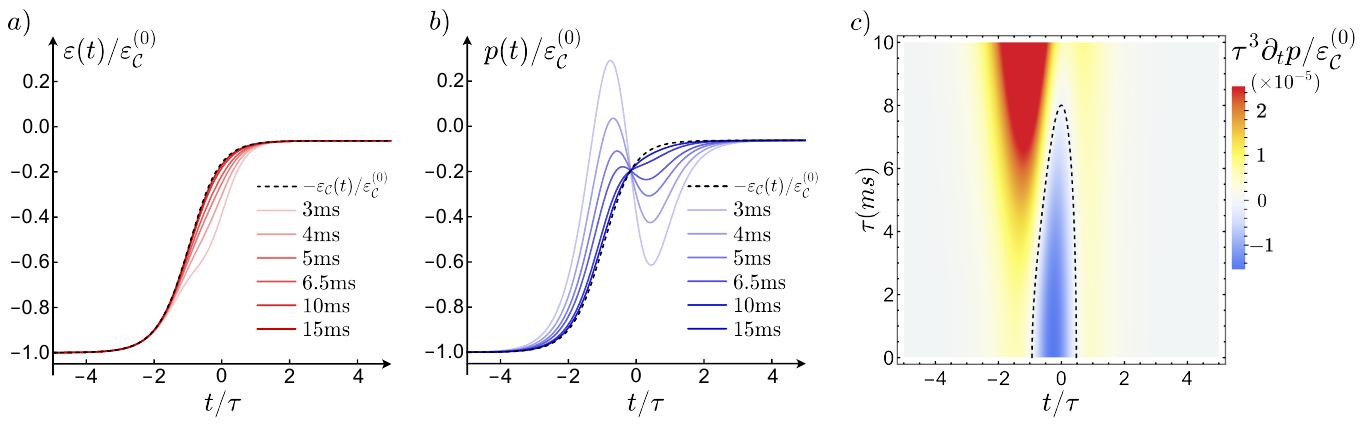}}
    \caption{{Anomalous thermodynamics  in an expanding ring.} 
    We consider the energy density $\varepsilon$ and radiative pressure $p$ of the vacuum along a ring. Its radius $R_0=10\mu$m undergoes a
    forced expansion over a time $\tau$ to a radius $R_1=4 R_0$.
    We set $c_s= 4.10^{-3}$m$s^{-1}$, similar to the sound velocity in a Bose-Einstein 
    $^{23}$Na 
    condensate \cite{eckel2018rapidly}. 
    The dashed line corresponds to $\varepsilon=p=-\varepsilon_{\mathcal{C}}(t)$ where $\varepsilon_{\mathcal{C}}(t)$ is the instantaneous Casimir energy density. 
We rescale these quantities by their initial value 
    $\varepsilon_{\mathcal{C}}^{(0)}=\varepsilon_{\mathcal{C}}(t=-\infty)$. 
    For a fast enough expansion
    with $\tau \leq 8$ms, an anomalous thermodynamics of the ring occurs. While 
    this does not manifest itself significantly on the evolution of the energy density a), the evolution of the radiative pressure is drastically altered and becomes non-monotonous b).  
    This anomalous regime occurs within the dashed region of panel c) where $\tau^3\partial_tp$ is plotted for various values of $\tau$.
    }
    \label{fig:expanding_BEC}
\end{figure*}

If the system size or the velocity is slowly varied in time,  
one  expects thermodynamic observables to be set by the instantaneous Casimir energy density 
$\varepsilon_{\mathcal{C}}(t)$ defined in \eqref{eq:CasimirEnergy} with $R(t)$. 
However, a new time-scale characterizing the expansion appears in the problem, 
$\partial_tR/R$. The associated energy-scale is set by the speed of variation of the size $R(t)$:
\begin{equation}
\label{eq:eps_sec}
    \varepsilon^{(a)}(t)= \frac{\hbar}{48\pi c_s}\left(\frac{\partial_tR}{R}\right)^2
\end{equation}
As a consequence we expect an energy density given by their sum: 
\begin{equation}
    \label{eq:energy}
    \varepsilon=-\frac{\mathcal{C}_w}{2}\left(\varepsilon_\mathcal{C}+\varepsilon^{(a)}\right)=-\frac{ \mathcal{C}_w}{2} \varepsilon_\mathcal{C} \left(
    1+\left(\frac{\partial_tR}{c_s}\right)^2\right)\,.
\end{equation}
As we will show later, it can also be inferred more rigorously from the modifications of the 
vacuum properties induced by a non-zero spacetime curvature known as gravitational 
anomalies~\cite{capper1974one,capper1974photon,bertlmann2000anomalies}.
As a consequence, when the radial velocity $\partial_tR$ becomes comparable with the longitudinal 
sound velocity $c_s$, we expect the instantaneous Casimir energy density to be modified by the expansion.

Similarly, we expect the radiative pressure to be altered by a fast expansion. 
Assuming a fast local equilibration, this radiative pressure satisfies
\begin{equation}
    \label{eq:thermo}
    p=-\frac{\partial_tE}{\partial_tR}
\end{equation}
relating the total energy in the system $E=2\pi R\varepsilon$ to the radiative pressure, such that 
\begin{equation}
    \label{eq:pressure}
    p=-\frac{ \mathcal{C}_w}{2} \varepsilon_\mathcal{C} \left(
    1+\left(\frac{\partial_tR}{c_s}\right)^2-2\frac{R \partial_t^2R}{c_s^2}\right)\,. 
\end{equation}
Moreover, we notice that contrarily to the static case, the radiative pressure and the energy 
density differ: $\varepsilon-p=\varepsilon^{(b)}=- \mathcal{C}_w \frac{\hbar}{24\pi c_s}\frac{\partial_t^2 R}{R}$.  
This manifest the breaking of the scale symmetry by the emergent energy scales $\varepsilon^{(a)}$ 
and $\varepsilon^{(b)}$.

To illustrate the experimental relevance of these corrections, we evaluate their 
amplitude in the context of an expanding Bose-Einstein condensate’s ring 
corresponding to $\mathcal{C}_w=1,\mathcal{C}_g=0$ (see the SM~\cite{SuppMat}). 
We consider a fast expansion of the ring's radius  from $R_0$ 
to $R_1$ over a time $\tau$, as shown in Fig.~\ref{fig:expanse}. 
Although our approach applies to a generic radius profile $R(t)$, for the sake of clarity we discuss the results for the profile 
\begin{equation}
    R(t)=\frac{R_0+R_1}{2}+\frac{R_1-R_0}{2}\tanh\left(\frac{t}{\tau}\right)\,.
\end{equation}
The parameters of Fig.~\ref{fig:expanding_BEC} are motivated by the experimental work of 
S.~Eckel \textit{et al.} on a $^{23}$Na 
Bose-Einstein condensate described 
in~\cite{eckel2018rapidly}. 
For a slow expansion of the condensate, represented in Fig.~\ref{fig:expanding_BEC}(a) for 
$\tau \geq 8$ms, the ratio $\partial_tR/c_s$ and thus 
$\varepsilon^{(a)}$ are negligible. As a consequence,  
a single anomalous energy scale 
$ \varepsilon^{(b)}$ controls the correction to the thermodynamics. 
The energy density is well approximated by the instantaneous Casimir energy 
$\varepsilon \approx -\frac12 \mathcal{C}_w \varepsilon_{\mathcal{C}}(t)$
while the pressure is corrected  to 
$p\approx- \frac{\mathcal{C}_w
}{2}\varepsilon_{\mathcal{C}}(t)-\mathcal{C}_w\varepsilon^{(b)}$.\\
For faster expansions of the condensate both energy scales $\varepsilon_{\mathcal{R}}$ and $\varepsilon_{\Bar{\mathcal{R}}}$ are sizable. 
While a small time dependent correction of the energy density 
appears as shown in Fig.~\ref{fig:expanding_BEC}(a), 
the more striking signature of the anomalous {dynamical} Casimir effect concerns the evolution of the pressure, represented in Fig.~\ref{fig:expanding_BEC}(b), which becomes non-monotonous. Such an oscillatory pressure is characteristic of an instability of the linear modes, which we expect to manifest itself by the excitation of higher energy modes. 

\begin{figure}
    \centering
    \includegraphics[width=\columnwidth]{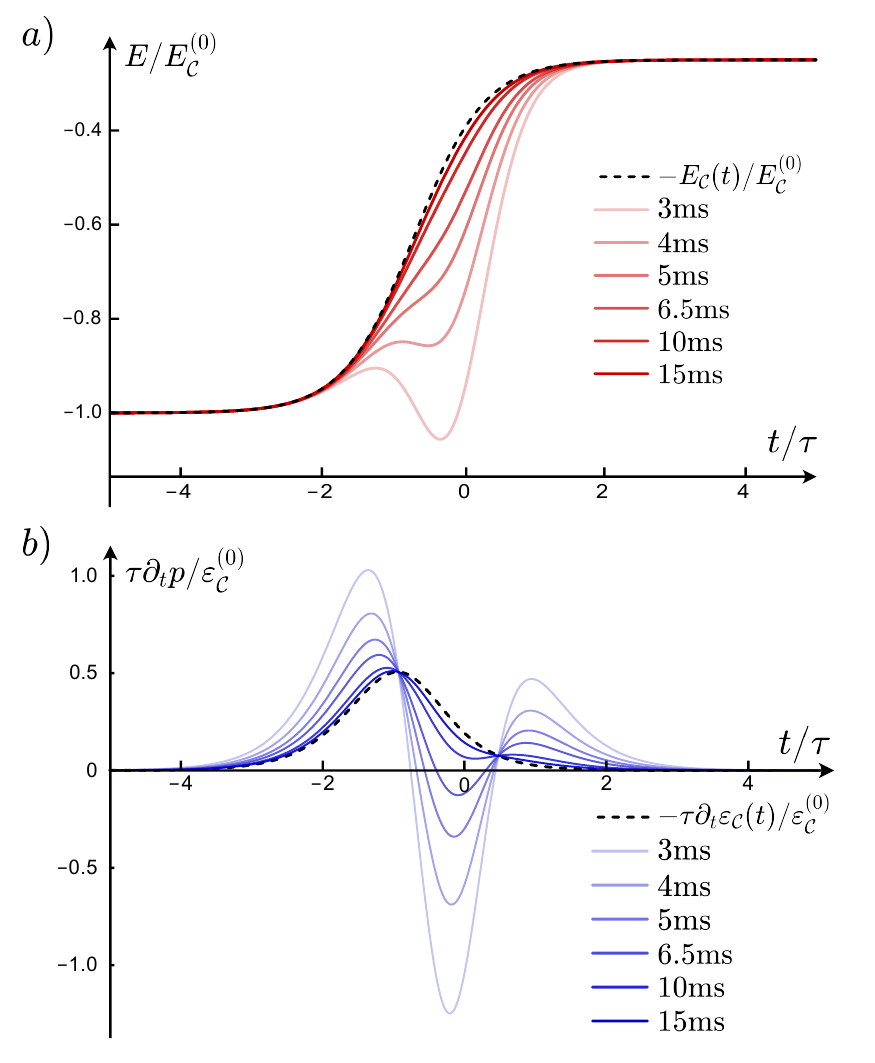}
    \caption{Energy of the driven expanding  ring.
    a) Vacuum energy of the ring. The dashed line corresponds to the instantaneous Casimir energy of the ring upon an increase of its size. The departure from this Casimir energy is a signature of the dynamical quantum effect predicted in this paper. While the total work needed for the expansion is independent of the rapidity of the drive, the profile of the energy injected into the ring depends on the duration $\tau$ of the ring's size increase, manifesting the effect of the conformal anomaly in this proposed experiment. 
    b) Evolution of the time-derivative of the radiative pressure. A mechanical instability of the ring, at which the vacuum's compressibility diverges, occurs when $\partial_t p=0$.
    }
    \label{fig:Energy}
\end{figure}

The stability condition can be inferred from the compressibility 
$\beta=-V^{-1} \partial V / \partial p = -R^{-1}~ \partial_tR / \partial_t p $. 
An instability develops whenever $\partial_tp$ vanishes, which occurs for fast enough expansions
of the ring as shown in Fig.~\ref{fig:Energy}(b) and in a rescaled manner in Fig.~\ref{fig:expanding_BEC}(c) for various $\tau$. It appears that the instability arises for fast enough expansions of the ring satisfying  $2c_s\tau<\sqrt{R_0^2+R_1^2+6R_0R_1}$.
In particular, if $c_s\tau<\sqrt{2}R_0$ they arise for any amplitude $R_1/R_0$ of the expansion. Note that in our case, the compressibility of the linear modes
is always negative, a consequence of the Casimir effect on a finite size system~\cite{cho2022quantum}. 
The presence of an instability, as well as the difference between energy and pressure, also have consequences on the external work required to drive the 
expansion of the ring. 
The work is deduced from the total energy of the vacuum along the ring which reads 
\begin{equation}
    E(t)=2\pi R(t)\varepsilon(t)
    =-E_{\mathcal{C}}(t)\left(1+\left(\frac{\partial_t R(t)}{c_s}\right)^2\right)\,.
\label{eq:EnergyTot}
\end{equation}

The resulting $E(t)$ clearly departs from the instantaneous Casimir energy 
$E_{\mathcal{C}}(t)=
\hbar c_s \mathcal{C}_w  / (24  R(t))$, as shown in Fig.~\ref{fig:Energy}(a). 
This departure, which corresponds to the change in the work necessary to drive the radius of the ring, is a 
signature of the anomalous {dynamical} Casimir effect discussed in this paper. \\

The emergence of the new  energy scales $\varepsilon^{(a)}$ 
and $\varepsilon^{(b)}$ and their effect on out-of-equilibrium thermodynamics quantities can be derived
from the modified vacuum properties of the associated quantum field theory, in the presence of a curved spacetime.
These modifications are captured by the conformal and gravitational anomalies: once quantized in a curved spacetime with a scalar curvature $\mathcal{R}\neq0$, the massless theory ceases to 
conserve 
the vacuum expectation value of the trace and covariant derivative of the stress-energy tensor $\mathcal{T}^{\mu\nu}$ \cite{capper1974one,capper1974photon,bertlmann2000anomalies}:

\begin{subequations}
\label{non-conservation}
	\begin{align}
 \label{non-conservation1}
    \mathcal{T}^\mu_{\phantom\mu\mu} &=\frac{\hbar c_s }{48\pi}\mathcal{C}_w\mathcal{R},  \qquad 
    \mathcal{T}^{\mu\nu} =\mathcal{T}^{\nu\mu}\,,\\
    \label{non-conservation2}
    \nabla_\mu\mathcal{T}^{\mu\nu} &=\frac{\hbar c_s}{96\pi}\mathcal{C}_g\frac{\epsilon^{\nu\mu}}{\sqrt{-g}}\nabla_\mu\mathcal{R}\,, 
	\end{align}
\end{subequations}
with $\mathcal{C}_g=c_+ - c_-$. 
The solutions of equations \eqref{non-conservation} for  a dynamical metric of the form \eqref{eq:metric}
can be decomposed as a sum (see the SM~\cite{SuppMat})
$\mathcal{T}=\mathcal{T}_{cl}+\mathcal{T}_q$. 
$\mathcal{T}_{cl}$ is the classical momentum energy tensor which satisfies the classical equation of motion 
\eqref{non-conservation} with $\mathcal{R}=0$. 
For an homogeneous initial vacuum state, it satisfies 
\begin{equation}
\label{eq:Tcl}
    \left[\mathcal{T}_{cl}(t)\right]^\mu_{\phantom{\mu}\nu}=
    -\frac{1}{2}\begin{pmatrix}
        \mathcal{C}_w&\mathcal{C}_g \frac{R_0}{R}\\
        -\mathcal{C}_g \frac{R}{R_0} &-\mathcal{C}_w
    \end{pmatrix}\varepsilon_{\mathcal{C}}(t)\,,
\end{equation}
characterized by the instantaneous Casimir energy density $\varepsilon_{\mathcal{C}}(t)=\hbar c_s / (24\pi R^2(t))$. 
This classical contribution is corrected by the quantum correction
\begin{equation}
\label{eq:anomalous_correction}
    \left[\mathcal{T}_{q}(t)\right]^\mu_{\phantom{\mu}\nu}=
    \frac12 
    \begin{pmatrix}
        \mathcal{C}_w \varepsilon_{\Bar{\mathcal{R}}} 
        & \mathcal{C}_g \frac{R_0}{R}(\varepsilon_{\Bar{\mathcal{R}}} - \varepsilon_\mathcal{R})\\
        \mathcal{C}_g\frac{R}{R_0}( \varepsilon_\mathcal{R} - \varepsilon_{\Bar{\mathcal{R}}} )
        & \mathcal{C}_w \left(2 \varepsilon_\mathcal{R}-\varepsilon_{\Bar{\mathcal{R}}} \right)
    \end{pmatrix}\,,
\end{equation}
whose amplitude is set by the new energy scales originating from the gravitational anomalies
\begin{equation}
\label{eq:epsR}
    \varepsilon_\mathcal{R} (t)= \frac{\hbar c_s}{48\pi}\mathcal{R}\,, \quad 
    \mathcal{R} = - \frac{2 }{ c^2_s }\frac{\partial_t^2 R}{R} \,.
\end{equation}
fixed by the spacetime scalar curvature $\mathcal{R}$ and 
\begin{equation}
\label{eq:epsRbar}
    \varepsilon_{\Bar{\mathcal{R}}}(t)=\frac{\hbar c_s}{48\pi}\Bar{\mathcal{R}}\,, \quad 
    \Bar{\mathcal{R}}= -2 c_s^{-2}\left(\frac{\partial_tR}{R}\right)^2\,.
\end{equation}
fixed by its time average $ \Bar{\mathcal{R}}=1/R^2\int\mathcal{R}\partial_tR$.
While $\varepsilon_\mathcal{R}$ controls the energy-pressure anisotropy $\varepsilon-p=\mathcal{C}_w\varepsilon_\mathcal{R}$, 
both new energy scales set the corrections to the Casimir energy density 
$\varepsilon+p=\mathcal{C}_w (\varepsilon_{\mathcal{C}}- \varepsilon_\mathcal{R}+\varepsilon_{\Bar{\mathcal{R}}} )$.
{
Note that the diagonal part of the tensor \eqref{eq:anomalous_correction}, related to the conformal anomaly, was identified in \cite{davies1977quantum, xie2023optomechanical}. However in these references, the authors focused on the evolution of the energy density in the ring, but did not discuss the behavior of the pressure and the associated instability, which are the core of the physics discussed here.}

These corrections to the momentum energy tensor induced by the time variations of the metric appear similar to those 
obtained in another context when considering inhomogeneous stationary metrics either close to a black hole~\cite{christensen1977trace}, or in a wire with an inhomogeneous temperature~\cite{bermond2022anomalous}. 
However, in our case, the variation of the metric in time and not in space leads to an change of the corrections to 
the energy density $\varepsilon$ and to the pressure $p$, and thus to new physical phenomena which we explore earlier in this paper. 

From Eqs.~(\ref{eq:Tcl},\ref{eq:anomalous_correction}), we get the expression of the energy and momentum density, the pressure, and the energy current of the azimuthal modes as 
\begin{subequations}
    \label{eq:thermo_1st_senario}
\begin{align}
\label{eq:e_scenario1}
    \varepsilon &= -\frac{ \mathcal{C}_w}{2} \varepsilon_\mathcal{C} \left(
    1+\left(\frac{\partial_tR}{c_s}\right)^2\right)\,,\\
    p &= \varepsilon + \mathcal{C}_w \varepsilon_\mathcal{C} ~ \frac{R \partial_t^2R}{c_s^2} \, , 
    \label{eq:p_scenario1}\\
    J_\varepsilon &=c_s^2\Pi=- \frac{\mathcal{C}_g }{2} \varepsilon_\mathcal{C}  c_s
     \left(
     1 +  \left(\frac{\partial_tR}{c_s}\right)^2
     -\frac{R \partial_t^2R}{c_s^2}
      \right)\,.
\label{eq:EnergyCurrent}
\end{align}
\end{subequations}
The  instantaneous Casimir energy density 
$\varepsilon_{\mathcal{C}}(t)$, defined in \eqref{eq:CasimirEnergy}, is therefore modified when either the radial velocity $\partial_tR$ is comparable to the particle velocity $c_s$, 
or the acceleration $\partial_t^2R$ is comparable to $c^2_s/R$. A comparison 
between (\ref{eq:epsR},\ref{eq:epsRbar}) and \eqref{eq:eps_sec}
reveals the identity $\varepsilon_\mathcal{R}\equiv\varepsilon^{(b)}$ and $\varepsilon_{\Bar{\mathcal{R}}}\equiv\varepsilon^{(a)}$. Finally, had we considered a chiral quantum ring, defined as one where $c_+\neq c_-$ corresponding to $\mathcal{C}_g\neq 0$, 
the current conservation equation will be modified into Eq.~\eqref{non-conservation2} as a consequence of the gravitational anomalies~\cite{bermond2022anomalous}. In this context, the anomalous {dynamical} Casimir effect manifests itself as a non-monotonous evolution of the energy current as the size is increased, as inferred from Eq.~\eqref{eq:EnergyCurrent}. 
\\

In this work, we have described a new effect, the anomalous {dynamical} Casimir effect, that alters the groundstate of a massless theory or vacuum 
in an expanding ring geometry. 
Because of the anomalous {dynamical} Casimir effect the required work profile to implement this expansion becomes non-monotonous for fast drives.
Such a profile is required to counteract the mechanical instability of the ring manifested by a diverging compressibility during the expansion. 
The modification of the thermodynamics of the vacuum is, in the language of quantum field theory, due to the conformal and gravitational anomalies. 
In the SM~\cite{SuppMat} we have also shown that a similar instability occurs then the speed of excitations $c_s$ is varied over time, an alternative protocol to varying the ring's radius as described in the main text. 
We have shown that the anomalous {dynamical} Casimir effect is distinct from the dynamical Casimir effect: it 
originates from the non-conservation equations induced by the conformal and gravitational anomalies~\eqref{non-conservation}, in contrast to the momentum energy tensor of the dynamical Casimir effect, which satisfies the standard conservation equations.
The possibility to relate this anomalous {dynamical} Casimir effect to the origin of dark energy along the lines of \cite{Mottola_2011}, and to the effects of transplanckian physics on the expansion dynamics along the lines of~\cite{weinfurtner2007trans,cha2017probing} are 
fascinating open questions. 

This new phenomenon can be experimentally tested, {\it e.g.} in Bose-Einstein condensates, or in quantum circuits realized at the edge of quantum Hall samples. 
Indeed, while the experiment by Eckel \textit{et al} 
\cite{eckel2018rapidly} considered a 
condensate's ring submitted to a free expansion and not to a forced one, 
two experimental manifestations appear in agreement with the instability described in the present paper: first, oscillations of the condensate radius over time are a possible manifestation of the variation of the radiative pressure and the associated instability. 
Second, the instability can also 
be related to the development of vortices observed at later times.

Finally, let us mention a few limits of our approach. Our analysis is based on the thermodynamics of the vacuum state. Studying the evolution of the system beyond this regime is a stimulating perspective, see  {\it e.g.} \cite{carusotto2010density}. 
Moreover, in this work, we described the system using local thermodynamic quantities, following the assumption within the general hydrodynamics framework of a fast local equilibration~\cite{castro2016emergent,essler2016quench,bernard2016conformal}. In other words, we are assuming that the local equilibration time is much smaller than any other time scale of the problem, such as $R/\partial_tR$ and $R/c_s$. 

Relating our results to Bose-Einstein condensates requires extra care. Indeed, 
hydrodynamic equations in a quantum fluid identify with the equation of motion of 
quantum fields in a curved spacetime following two crucial assumtions. 
First, as described in the supplemental materials~\cite{SuppMat}, the dispersion relation can only be approximated as linear up to the healing wavelength of the condensate 
and second,  we need to 
neglect the time dependence of the speed of sound during the evolution~\cite{eckel2018rapidly}. 
It is possible to fulfill both assumptions experimentally. 
First, if we focus on ground state properties of the system, 
a description based on the linear dispersive low-energy modes appears sufficient unless instabilities excite higher energy modes. 
Then, while in any experimentally relevant protocol the speed of sound is non-constant due to the change in the background condensed density $n_0$ scaling following~\cite{eckel2018rapidly} as $c_s\propto R^{-2/7}$, it is possible to counterbalance this effect by modulating the interaction length using {\it e.g.} Feshbach resonances~\cite{vogels1999feshbach,inouye1998observation}.
Additionally, a full description of the problem would also require an extension of our work by incorporating the weak coupling between the 1D azimutal dynamics and the other three-dimensional degrees of freedom.  \\
\begin{acknowledgments}
We thank Maxim Chernodub and Sergio Ciliberto for stimulating discussions at various stages of this project. This work was supported by the French ANR grant PROCURPHY.
\end{acknowledgments}
\bibliographystyle{apsrev4-1}
\bibliography{Biblio}
\newpage
\setcounter{equation}{0}
\setcounter{figure}{0}
\setcounter{table}{0}
\setcounter{page}{1}
\onecolumngrid
\makeatletter
\renewcommand{\theequation}{S\arabic{equation}}
\renewcommand{\thefigure}{S\arabic{figure}}
\renewcommand{\bibnumfmt}[1]{[S#1]}
\renewcommand{\citenumfont}[1]{S#1}
\begin{center}
\textbf{\large SUPPLEMENTAL MATERIAL\\
Anomalous {dynamical} Casimir effect in an expanding ring}
\end{center}
\begin{bibunit}  
\section{Curved spacetimes in Bose-Einstein condensates}
In this appendix, following the review of Barcelo, Liberati, and Visser~\cite{barcelo2001analogue} we analyze the possibility to generate analog curved spacetimes in Bose-Einstein condensates.
    
\subsection{From Bose-Einstein condensates to hydrodynamics-like equations}

In a quantum system of $N$ interacting bosons, Bose-Einstein condensation corresponds to a configuration where a finite fraction of the bosons lie in the same single-particle quantum state. The observables of such a condensate can be deduced from a second quantized Hamiltonian of the form
\begin{equation}
        \mathcal{H}=\int\diff \Vec{x}\,\Hat{\Psi}\dag(\Vec{x},t)\left[-\frac{\hbar^2}{2m}\Vec{\nabla}^2+V_{ext}(\Vec{x})\right]\Hat{\Psi}(\Vec{x},t)
        +\frac12\int\diff \Vec{x}\,\diff\Vec{y}\,\Hat{\Psi}\dag(\Vec{x},t)\Hat{\Psi}\dag(\Vec{y},t)V\left(\Vec{x}-\Vec{y}\right)\Hat{\Psi}(\Vec{y},t)\Hat{\Psi}(\Vec{x},t)\,,
\end{equation}
with $V_{ext}(\Vec{x})$ an external trapping potential and $V(\Vec{x}-\Vec{y})$ the interatomic two-body interacting potential. 
When $N$ is large, computing the spectrum of this Hamiltonian and solving for $\Hat{\Psi}$ becomes impractical. 
We resort to the mean-field  Bogoliubov ansatz 
\begin{equation}
    \Hat{\Psi}(\Vec{x},t)=\psi(\Vec{x},t)+\Hat{\phi}(\Vec{x},t)\,,
\end{equation}
where $\psi(\Vec{x},t)=\langle\Hat{\Psi}(\Vec{x},t)\rangle$ is referred to as the wave function 
of the condensate~\cite{bogoliubov1947theory}, while $\Hat{\phi}$ denotes the excitation on top of the condensate.\\

Assuming that the number of particles that do not lie in the condensate wavefunction is small compared to $N$, a zeroth order approximation in the non condensate number of particle
leads to the equation of motion of the condensate
\begin{equation}
    i\hbar\partial_t\psi(\Vec{x},t)=\left[-\frac{\hbar^2}{2m}\Vec{\nabla}^2+V_{ext}(\Vec{x})+\int\diff\Vec{y}\,\psi\dag(\Vec{y},t)V(\Vec{x}-\Vec{y})\psi(\Vec{y},t)\,\right]\psi(\Vec{x},t)\,.
\end{equation}
If we further assume the two-body interaction to be short range $V(\Vec{x})\approx\kappa\delta(\Vec{x})$ with $\kappa=\frac{4\pi a \hbar^2}{m}$ where $a$ represents the scattering length, the condensate wave function verifies the Gross-Pitaevskii (also known as the non-linear Shrödinger or time-dependent Landau-Ginzburg) equation
\begin{equation}
    \label{eq:Gross_Pitaevskii}
    i\hbar\partial_t\psi=\left(-\frac{\hbar^2}{2m}\Vec{\nabla}^2+V_{ext}(\Vec{x})+\kappa\psi\dag\psi\right)\psi\,,
\end{equation}
where for brevity we omitted the $(\Vec{x},t)$ dependence of the wave function.\\

We parameterize $\psi$ using the so-called Madelung representation with a phase and a density such as $\psi=\sqrt{n_0}\exp\left(i\theta_0/\hbar\right)$. 
The equations of motion for the phase $\theta_0$ and the density $n_0$ are given by
\begin{equation}
    \label{eq:Mean_field_background}
    \begin{cases}
        \partial_t n_0+\frac{1}{m}\Vec{\nabla}\cdot\left(n_0\Vec{\nabla}\theta_0\right)=0\,,\\
        \partial_t\theta_0+\frac{1}{2m}\left(\Vec{\nabla}\theta_0\right)\cdot\left(\Vec{\nabla}\theta_0\right)+V_{ext}+\kappa n_0-\frac{\hbar^2}{2m}\frac{\Vec{\nabla}^2\sqrt{n_0}}{\sqrt{n_0}}=0\,.
    \end{cases}
\end{equation}
They are identical to Euler equations for a barotropic inviscid irrotational fluid~\cite{unruh1981experimental} of velocity $\Vec{v}$ with $\Vec{v}\equiv \frac{1}{m}\Vec{\nabla}\theta_0$ in the presence of a quantum potential term $V_{quant}=-\frac{\hbar^2}{2m}\frac{\Vec{\nabla}^2\sqrt{n_0}}{\sqrt{n_0}}$.

\subsection{Sound propagation and metric tensor}
Starting from equation~\eqref{eq:Gross_Pitaevskii}, let us show that in a way similar to the classical hydrodynamic, studied by Unruh~\cite{unruh1981experimental}, the weak fluctuation of $\psi$ around a background $\psi_0$ can be described by a curved spacetime Klein-Gordon equation. To do so, let us develop $\psi$ around a background $\psi_0=\sqrt{n_0}\exp\left(i\theta_0/\hbar\right)$ such as
\begin{equation}
    \begin{cases}
        n=n_0+\delta n\,,\\
        \theta=\theta_0+\delta \theta\,.
    \end{cases}
\end{equation}
Assuming the perturbation to be small, there is no backreaction of the fluctuation $\delta n,\delta\theta$ on the background equations~\eqref{eq:Mean_field_background}, and the fluctuations verify an equation similar to a curved spacetime massless Klein Gordon field:
\begin{equation}
\label{eq:quantum_fluctuation}
    \begin{cases}
        \partial_t\delta n+\frac{1}{m}\Vec{\nabla}\cdot\left(\delta n\Vec{\nabla}\theta_0+n_0\Vec{\nabla}\delta \theta\right)=0\,,\\
        \partial_t\delta \theta+\frac{1}{m}\left(\Vec{\nabla}\delta \theta\right)\cdot\left(\Vec{\nabla}\theta_0\right)+\kappa \delta n-\frac{\hbar^2}{2m}D_2\left[\delta n\right]=0\,,
    \end{cases}
\end{equation}
with
\begin{equation}
    D_2\left[\delta n\right]=\frac{\Vec{\nabla}^2\delta n}{2n_0}-\frac{\Vec{\nabla}^2n_0}{2n_0}\delta n-\frac{\left(\Vec{\nabla}\delta n\right)\cdot\left(\Vec{\nabla}n_0\right)}{2n_0^2}+\frac{\left(\Vec{\nabla}n_0\right)^2}{2n_0^3}\delta n\,.
\end{equation}
Comparing with the result for classical hydrodynamic equation shows that a quantum correction appears through the operator $D_2$. However, on wavelengths larger than the healing length $\frac{2\pi}{k}>\xi=\frac{\hbar}{\sqrt{m\kappa n}}$, this term becomes negligible such that the equation of motion for $\delta\theta$~\eqref{eq:quantum_fluctuation} can be reduced to
\begin{equation}
    \partial_t\left(\frac{1}{\kappa}\partial_t\left(\delta\theta\right)\right)+\partial_t\left(\frac{\Vec{v}}{\kappa }\cdot\left(\Vec{\nabla}\delta\theta\right)\right)+\Vec{\nabla}\cdot\left(\frac{\Vec{v}}{\kappa }\partial_t\delta\theta\right)+\nabla_i\left(\frac{v_iv_j}{\kappa}\nabla_j\left(\delta\theta\right)\right)-\Vec{\nabla}\left(\frac{n_0}{m}\Vec{\nabla}\left(\delta\theta\right)\right)=0
\end{equation}
with $v_i=\frac{\partial_i\theta_0}{m}$ the fluid velocity. Rewriting this equation of motion as
\begin{equation}
\label{eq:f_curved}
    \partial_\mu\left(f^{\mu\nu}\partial_\nu\left(\delta\theta\right)\right)=0\,,
\end{equation}
with
\begin{equation}
    \label{eq:BEC_metric_0}
    f^{\mu\nu}=
    \begin{pmatrix}
        \frac{1}{\kappa}&\frac{v_i}{\kappa}\\
        \frac{v_j}{\kappa}& \frac{v_iv_j}{\kappa}-\frac{n_0}{m}\delta_{ij}
    \end{pmatrix}\,,
\end{equation}
one can interpret it, as the equation of motion for a massless scalar field in a curved spacetime whose metric $g_{\mu\nu}$ verifies
\begin{equation}
    \label{eq:BEC_metric}
    f^{\mu\nu}=\sqrt{-\det(g_{\rho\sigma})}g^{\mu\nu}=\frac{n_0}{mc_s^2}
    \begin{pmatrix}
        1&v_i\\
        v_j& -c_s^2\delta^{ij}+v_iv_j
    \end{pmatrix}\,,
\end{equation}
with $c_s^{2}=\frac{\kappa n_0}{m}$ the sound velocity. Note at this stage that the metric is dimensionful tensor. 
A dimensionless metric is obtained by factoring out a parameter set by a static reference value for the density and the velocities.\\

The precise form of the metric depends on the dimensionality of the system. 
In dimension $D+1$ with $D>1$, $\det\left(f^{\mu\nu}\right)=\det\left(g^{\mu\nu}\right)^{\frac{1-D}{2}}$, we can deduce the metric from equation~\eqref{eq:BEC_metric}:\begin{equation}
        \label{eq:App_A_metric}
        g_{\mu\nu}=\left(\frac{n_0}{m c_s^2}\right)^{\frac{2}{D-1}}\begin{pmatrix}
        c_s^2-\left|\Vec{v}\right|^2&v_i\\
        v_j& -\delta^{ij}
    \end{pmatrix}\,.
    \end{equation} 
In 1+1 dimensions, due to the conformal symmetry, every metric verifies 
$\det\left(\sqrt{-\det(g_{\rho\sigma})}g^{\mu\nu}\right)=1$. 
Therefore, whenever $\det\left(f^{\mu\nu}\right)$ is not a constant \eqref{eq:f_curved} does not describe the dynamics of a massless Boson in curved spacetime.
Starting from a metric $f^{\mu\nu}$ with $\det\left(f^{\mu\nu}\right)=Cst$, 
we identify a family of metrics $g^{\mu\nu}$ of the form 
$g_{\mu\nu}=e^{\sigma(x)}f_\mu\nu$ that satisfy 
\begin{equation}
    \sqrt{-\det(g_{\rho\sigma})}g^{\mu\nu}=f^{\mu\nu}\,.
\end{equation}
To determine the value of $g_{\mu\nu}$, a supplementary condition is necessary. Depending on the protocol, two strategies can be used:
\begin{itemize}
    \item In an experiment, every Bose-Einstein condensate is a 3+1 dimensional system. Therefore, a 1+1 dimensional dynamics is solely an approximation in which the dynamics is supposed frozen or slower in two of the dimensions, let's say y and z. The corresponding 1+1 dimensional metric is then the dimensional reduction of the 3+1 dimensional one obtained 
    by fixing $\diff y=\diff z=0$.
    \item Another strategy consists in identifying a thermodynamic quantity whose value depends on the metric. For example, let us consider a Bose-Einstein condensate with a sound velocity $c_s$ within a trap of size $L$ whose size and velocity are modulated in time. In the adiabatic limit, the energy density of such a system is given by the instantaneous Casimir energy
    \begin{equation}
    \label{eq:instantaneous_casimir_adiabatic}
        \varepsilon=-\frac{\pi\hbar c_s(t)}{6L^2(t)}\, .
    \end{equation}
    Since the adiabatic energy density, for a metric of the form 
    \begin{equation}
        g_{\mu\nu}=\begin{pmatrix}
            f_1(t)&0\\0&f_2(t)
        \end{pmatrix}
    \end{equation}
    is given by $\varepsilon=-\frac{\pi\hbar c_s^0}{6L^2_0}\frac{1}{f_2}$, direct comparison with \eqref{eq:instantaneous_casimir_adiabatic} can be used to deduced the value of $f_2(t)$, from which, together with \eqref{eq:BEC_metric_0} the value of $f_1$ can be deduced.
\end{itemize}

\subsection{Limit of the description and healing length}
Let us explore the limits of the description used in this appendix. In the limits of small wavelengths, the eikonal approximation applies. In this approximation, the amplitude of the fluctuations of the background are assumed to be small compared to the fluctuation, such that $D_2\left[\delta n\right]\approx\frac{1}{2n_0}\Vec{\nabla}^2\delta n$. Under this approximation, the energy spectrum becomes 
\begin{equation}
    \omega=\Vec{v}\cdot\Vec{k}\pm c_s\left|\Vec{k}\right|\to\omega=\Vec{v}\cdot\Vec{k}\pm\sqrt{c_s^2\Vec{k}^2+\left(\frac{\hbar}{2m}\Vec{k}^2\right)^2}\,.
\end{equation}
We recover a linear spectrum for large wavelengths ($2\pi/k>\xi$). Therefore, the dynamic of the corresponding particles is well described by the relativistic physics picture presented previously. For smaller wavelengths ($2\pi/k<\xi$), quantum corrections induce non-linearities, and the relativistic picture breaks down.\\
\section{Anomalous stress-energy tensor in time-dependent metrics}    
    In 1977, Christensen and Fulling \cite{christensen1977trace}  computed the stationary momentum energy tensor for both a 3+1 and a 1+1 dimensional Schwarzschild black hole. In a previous article \cite{bermond2022anomalous}, we generalized their argument to determine the stationary momentum energy tensor of a chiral field in any background 1+1 dimensional static metric. Here, we reconsider these arguments in the case of a 1+1 dimensional dynamical metric of the form 
    \begin{equation}
        \label{eq:App_B_metric}
        \diff s^2=f_1(t)c_s^2\diff t^2-f_2(t)\diff x^2\,.
    \end{equation}
    \subsection{Gravitational anomalies}
    Focusing on a symmetric momentum energy tensor, the anomalies, encoding the violation of classical symmetries by quantum fluctuations in the presence of a background metric, can be written~\cite{capper1974one,capper1974photon,bertlmann2000anomalies} as 
    \begin{subequations}
    \begin{itemize}
        \item The trace anomaly:
            \begin{align}
            \label{eq:App_B_Trace}
            {\mathcal T}^{\mu\nu}&=\mathcal{C}_w\frac{\hbar c_s}{48\pi}\mathcal{R}\,,
            \end{align}
        \item The Einstein anomaly:
            \begin{align}
            \label{eq:App_B_conservation}
            \nabla_\mu {\mathcal T}^{\mu\nu}&=\mathcal{C}_g\frac{\hbar c_s}{96\pi}\frac{1}{\sqrt{\abs{\mathrm{det}\left(g_{\rho\sigma}\right)}}}\varepsilon^{\nu\mu}\nabla_\mu \mathcal{R}\,,
            \end{align}
    \end{itemize} 
    \end{subequations}
where $\mathcal{R}$ represents the Ricci scalar curvature, $\varepsilon^{\mu\nu}$ is the totally antisymmetric tensor with $\varepsilon^{01}=1$, while $\mathcal{C}_g$ and $\mathcal{C}_w$ are the gravitational anomaly coefficients, 
\begin{subequations}
\label{eq:App_B_coefficients}
\begin{align}
    \mathcal{C}_w&=\sum c\,,\\
    \mathcal{C}_g&=\sum \chi c\,,
\end{align} 
\end{subequations}
with $c$ the central charges and $\chi$ the chiralities. In other words, this corresponds to 
$\mathcal{C}_g = c_+ - c_-$ and $\mathcal{C}_w= c_+ + c_- $. 
A simple proof of the formula (\ref{eq:App_B_Trace},\ref{eq:App_B_conservation}) is derived by Bertlmann and Kohlprath in \cite{bertlmann2001two} for free chiral fermions with $\mathcal{C}_w=1$ and $\mathcal{C}_g=\pm1$. 
\subsection{Background metric properties}
As mentioned above, for convenience,  we will focus on the case of a diagonal, dynamical metric given by
\begin{equation}
\label{eq:App_B_coordinates}
    \diff s^2=f_1(t)c_s^2\diff t^2-f_2(t)\diff x^2.
\end{equation}
Even though in two dimensions and for any 1+1 dimensional manifold, there exist global coordinates in which the metric is of the form 
\begin{equation}
    \diff s^2=\Omega(x,t)^2\left(c_s^2\diff t^2-\diff x^2\right),
\end{equation}
we will use to the diagonal metric \eqref{eq:App_B_coordinates}, which is convenient to express the results in the original laboratory coordinates.\\

In the metric \eqref{eq:App_B_coordinates}, the non-zero Christoffel symbols 
\begin{equation}
    \label{eq:levi-civita}
     \begin{Bmatrix}\nu\\\rho\mu\end{Bmatrix}=\frac{1}{2}g^{\nu\sigma}\left(\partial_\rho g_{\sigma\mu}+\partial_\mu g_{\rho\sigma}-\partial_\rho g_{\rho\mu}\right),
\end{equation}
are
\begin{equation}
\label{eq:App_B_non_0_levi}
    \begin{aligned}
        \begin{Bmatrix}0\\00\end{Bmatrix}&=\frac{1}{2c_s}\frac{\partial_tf_1}{f_1}\,,\\
        \begin{Bmatrix}1\\01\end{Bmatrix}=\begin{Bmatrix}1\\10\end{Bmatrix}&=\frac{1}{2c_s}\frac{\partial_tf_2}{f_2}\,,\\
        \begin{Bmatrix}0\\11\end{Bmatrix}&=\frac{1}{2c_s}\frac{\partial_tf_2}{f_1}\,,
    \end{aligned}
\end{equation}
The corresponding non-zero Riemann tensor coefficients
\begin{align}
     \label{eq:non_0_riemann}
    \mathcal{R}^\mu_{\phantom{\mu}\nu\rho\sigma} =\partial_\rho\begin{Bmatrix}\mu\\\nu\sigma\end{Bmatrix}-\partial_\sigma\begin{Bmatrix}\mu\\\nu\rho\end{Bmatrix}
    +\begin{Bmatrix}\lambda\\\nu\sigma\end{Bmatrix}\begin{Bmatrix}\mu\\\lambda\rho\end{Bmatrix}-\begin{Bmatrix}\lambda\\\nu\rho\end{Bmatrix}\begin{Bmatrix}\mu\\\lambda\sigma\end{Bmatrix}
\end{align}
are 
\begin{eqnarray}
\label{eq:non_0_riemann_applied}
       \mathcal{R}^0_{\phantom{t}x0x}&=&-\mathcal{R}^0_{\phantom{t}xx0}\\
       \nonumber
       &=&\frac{1}{2c_s^2}\left(\frac{\partial_t^2f_2}{f_1}
       -\frac{1}{2}\frac{\partial_tf_2}{f_1}\left(\frac{\partial_tf_1}{f_1}+\frac{\partial_tf_2}{f_2}\right)
       \right),\\
       \mathcal{R}^x_{\phantom{t}0x0}&=&-\mathcal{R}^x_{\phantom{t}00x}\\
        \nonumber
       &=&-\frac{1}{2c_s^2}\left(\frac{\partial_t^2f_2}{f_2}
       -\frac{1}{2}\frac{\partial_tf_2}{f_2}\left(\frac{\partial_tf_1}{f_1}+\frac{\partial_tf_2}{f_2}\right)
       \right)\,. 
\end{eqnarray}
Hence the curvature Ricci scalar reads 
\begin{align}
        \mathcal{R} & =  g^{\rho\nu}\mathcal{R}^\mu_{\phantom{\mu}\nu\mu\rho} \nonumber \\
          &=\frac{1}{c_s^2}\left[-\frac{\partial_t^2f_2}{f_1f_2}+\frac{1}{2}\frac{\partial_tf_2}{f_1f_2}\left(\frac{\partial_tf_2}{f_2}+\frac{\partial_tf_1}{f_1}\right)\right]\,.
        \label{ricci_applied}
\end{align}
\subsection{Momentum energy tensor}
Let us now consider the (non-)conservation equation \eqref{eq:App_B_conservation} in this curved spacetime. 
We rewrite the two equations (for $\nu=0,x$) 
in the coordinates \eqref{eq:App_B_metric} as
\begin{align}
\label{eq:conservation_applied}
       \partial_0{\mathcal T}^0_{\phantom{x}0}+\partial_x{\mathcal T}^x_{\phantom{x}0}+\frac{1}{2c_s}\frac{\partial_tf_2}{f_2}\left({\mathcal T}^0_{\phantom{x}0}-{\mathcal T}^x_{\phantom{x}x}\right)&=0\\
\label{eq:conservation_applied2}
       \partial_0{\mathcal T}^0_{\phantom{x}x}+\partial_x{\mathcal T}^x_{\phantom{x}x}+\frac{1}{2c_s}\frac{\partial_tf_1}{f_1}{\mathcal T}^0_{\phantom{x}x}-\frac{1}{2c_s}\frac{\partial_xf_2}{f_1}{\mathcal T}^x_{\phantom{0}0}
       &=\mathcal{C}_g\frac{\hbar}{96\pi}\sqrt{\frac{f_2}{f_1}}\partial_t\mathcal{R},
\end{align}
Using the symmetry properties  ${\mathcal T}_{\mu\nu}={\mathcal T}_{\nu\mu}$, expressed as 
\begin{equation}
    {\mathcal T}^0_{\phantom{x}x}=-\frac{f_2}{f_1}{\mathcal T}_{\phantom{x}0}^x
\end{equation}
and, the trace anomalies in 1+1 dimension, relating the trace of the energy momentum tensor to the spacetime geometry for conformal
theories in 1+1 dimensions 
\begin{equation}
\label{eq:trace}
    {\mathcal T}^\alpha_{\phantom{\alpha}\alpha}={\mathcal T}^0_{\phantom{0}0}+{\mathcal T}^x_{\phantom{0}x}=\mathcal{C}_w\frac{\hbar c_s}{48\pi}\mathcal{R},
\end{equation}
we simplify (\ref{eq:conservation_applied},\ref{eq:conservation_applied2}) into  
\begin{subequations}
\label{eq:simplify_conservation}
    \begin{align}
        \partial_0\left[f_2{\mathcal T}^0_{\phantom{0}0}\right]-\partial_x\left[f_1{\mathcal T}^0_{\phantom{0}x}\right]&=\mathcal{C}_w\frac{\hbar}{96\pi}\mathcal{R}\partial_tf_2\,,\label{eq:energy}\\
        \partial_0\left[\sqrt{f_1f_2}{\mathcal T}^0_{\phantom{0}x}\right]-\partial_x\left[\sqrt{f_1f_2}\mathcal{T}^0_{\phantom{x}0}\right]&=\mathcal{C}_g\frac{\hbar}{96\pi}f_2\partial_t\mathcal{R}\,.\label{eq:current}
    \end{align}
\end{subequations}
Whose solutions are given by
\begin{equation}
    \label{eq:momentum_Energy_Tensor}
    {\mathcal T}^\mu_{\phantom{0}\nu}=\left(T^\mu_{\phantom{0}\nu}\right)_0+\left(T^\mu_{\phantom{0}\nu}\right)_{an}\,,
\end{equation}
with $\left(T^\mu_{\phantom{0}\nu}\right)_0$, the classical solution (verifying \eqref{eq:simplify_conservation} with $\mathcal{R}=0$)
\begin{equation}
    \left({\mathcal T}^\mu_{\phantom{0}\nu}\right)_{0}=
    \begin{pmatrix}
    F(x+c_s\tau)+G(x-c_s\tau)&\sqrt{\frac{f_1}{f_2}}\left(G(x-c_s\tau)-F(x+c_s\tau)\right)\\\sqrt{\frac{f_2}{f_1}}\left(F(x+c_s\tau)-G(x-c_s\tau)\right)&-F(x+c_s\tau)-G(x-c_s\tau)
    \end{pmatrix}\frac{1}{2f_2},
\end{equation}
specified by $F$ and $G$ two function to be fixed by boundary conditions, with $\diff\tau=\sqrt{\frac{f_1}{f_2}}\diff t$, and
\begin{equation}
\begin{aligned}
    \left({\mathcal T}^\mu_{\phantom{0}\nu}\right)_{an}&=
    \frac12\begin{pmatrix}
    \mathcal{C}_w\varepsilon_{\Bar{\mathcal{R}}}&\mathcal{C}_g\sqrt{\frac{f_1}{f_2}}\left(\varepsilon_{\Bar{\mathcal{R}}}-\varepsilon_\mathcal{R}\right)\\\mathcal{C}_g\sqrt{\frac{f_2}{f_1}}\left(\varepsilon_\mathcal{R}-\varepsilon_{\Bar{\mathcal{R}}}\right)&\mathcal{C}_w\left(2\varepsilon_\mathcal{R}-\varepsilon_{\Bar{\mathcal{R}}}\right)
    \end{pmatrix}\,.\\
\end{aligned} 
\end{equation}
with the new energy scales
\begin{equation}
\begin{aligned}   
    \varepsilon_\mathcal{R}&=\frac{\hbar c_s}{48\pi}\mathcal{R}\quad\textrm{and}\quad
    \varepsilon_{\Bar{\mathcal{R}}}&=\frac{\hbar c_s}{48\pi}\Bar{\mathcal{R}}\,,
\end{aligned}
\end{equation}
where
\begin{equation}
   \Bar{\mathcal{R}}=\frac{1}{f_2}\int\diff t\mathcal{R}\partial_tf_2=-\frac{1}{c_s^2}\frac{1}{2f_1}\left(\frac{\partial_tf_2}{f_2}\right)^2\,.
\end{equation}
\section{Anomalous {dynamical} Casimir effect for a Bose-Einstein condensate time-dependent velocity}
In the main text, we considered a situation in which one induces an effective curvature of spacetime in a Bose-Einstein condensate by modulating the trapping shape. Another experimentally relevant strategy to induce a non-zero spacetime curvature consists in modulating over time the sound velocity of the condensate. Such a variation can in principle be implemented by resorting to Feshbach resonances~\cite{vogels1999feshbach,inouye1998observation} and leads to a dynamical metric of the form 
\begin{equation}
    \diff s^2=\frac{c_s(t)}{c_0}c_0^2\diff t^2-\frac{c_0}{c_s(t)}\diff x^2\,.
\end{equation}
Therefore, in this section, we study the consequences of gravitational anomalies on the thermodynamic properties of such a velocity modulated Bose-Einstein condensates.\\

In such an velocity modulated Bose-Einstein ring, the new energy scales turn out to be defined 
as
\begin{align}
       \varepsilon_\mathcal{R}=\frac{\hbar}{48\pi c_s }\left[\frac{\partial_t^2c_s}{c_s}-2\left(\frac{\partial_tc_s}{c_s}\right)^2\right]\,,\\
       \varepsilon_{\Bar{\mathcal{R}}}=-\frac{\hbar}{96\pi c_s}\left(\frac{\partial_tc_s}{c_s}\right)^2\,,
\end{align}
such as the energy density, the pressure, and the energy current, in a system of size $L$ read
\begin{subequations} \label{eq:MET_scattering_bis}
\begin{align} 
        \varepsilon &=-\frac{\mathcal{C}_w}{2}\left(\varepsilon_\mathcal{C}(t)+\frac{\hbar}{96\pi c_s}\left(\frac{\partial_tc_s}{c_s}\right)^2\right)\,,\\
        p&=\varepsilon-\mathcal{C}_w\frac{\hbar}{48\pi c_s }\left[\frac{\partial_t^2c_s}{c_s}-2\left(\frac{\partial_tc_s}{c_s}\right)^2\right]\,\\
        J_\varepsilon &=-\frac{\mathcal{C}_g}{2}\left(-c_s\varepsilon_\mathcal{C}(t)+\frac{\hbar}{48\pi}\left[\frac{\partial_t^2c_s}{c_s}-\frac32\left(\frac{\partial_tc_s}{c_s}\right)^2\right]\right)\,.
\end{align}
\end{subequations}
The finite size instantaneous Casimir energy density $\varepsilon_{\mathcal{C}}(t)=\frac{\pi\hbar c_s(t)}{6L^2}$ is therefore modified by the presence of an external acceleration $\partial_tc_s$. In particular, the energy density is reminiscent of the result obtained by W.Unruh in a different context\cite{unruh1976notes}. Indeed, defining $T_U$ as the Unruh temperature such as
\begin{equation}
    k_BT_U=\frac{\hbar}{2\pi}\frac{\partial_tc_s}{c_s}
\end{equation}
with $\partial_tc_s$ the instantaneous acceleration, the anomalous correction to the energy density takes the form
\begin{equation}
    \delta\varepsilon=\frac{\mathcal{C}_w}{8}\frac{\pi}{6\hbar c_s}k_B^2T_U^2\,,
\end{equation}
wich corresponds to a quarter of Unruh thermal energy, $\varepsilon_U=\mathcal{C}_w\gamma T_U^2/2$~\cite{fulling1973nonuniqueness,davies1975scalar,unruh1976notes}. Therefore, the total energy density is the sum of the instantaneous Casimir energy and the instantaneous Unruh energy density.\\

\begin{figure}
    \centering
    \includegraphics[width=\textwidth]{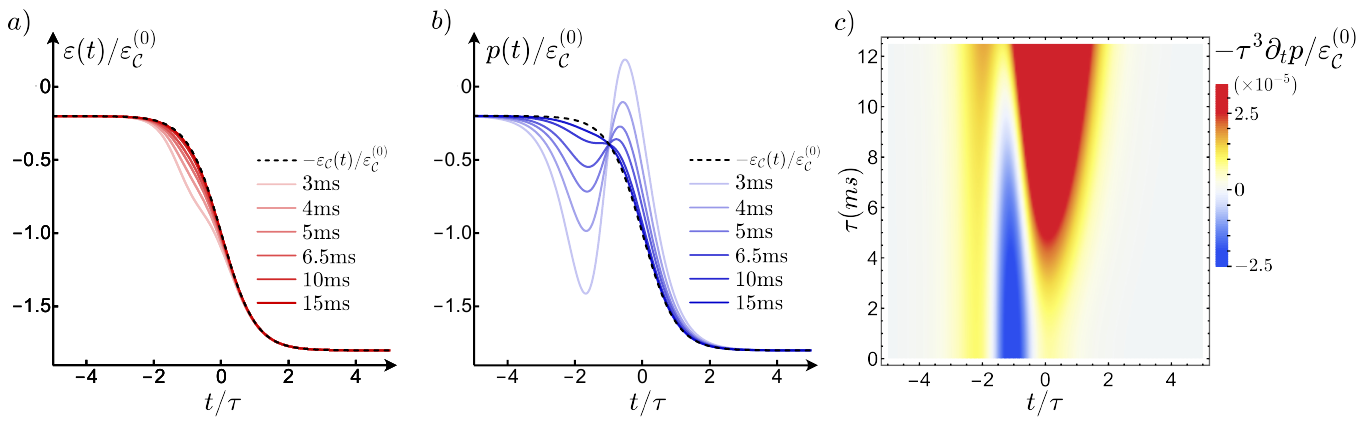}
    \caption{{Anomalous thermodynamics in a Bose-Einstein condensates with a time-dependent velocity}
    We consider the energy density $\varepsilon$ and radiative pressure $p$ of the vacuum along a Bose-Eistein condensate trapped in a finite size trap of length $L\approx10\mu$m. Its sound velocity $c_s^0$ undergoes an increase over a time $\tau$ to a sound velocity $c_s^1=9c_s^0$.
    We set $c_s^0=8.10^{-4}$m$s^{-1}$, such that the velocities $c_s^0$ and $c_s^1$ involved are similar to the sound velocity in a Bose-Einstein 
    $^{23}$Na 
    condensate \cite{eckel2018rapidly}. 
    The dashed line corresponds to $\varepsilon=p=-\varepsilon_{\mathcal{C}}(t)$ where $\varepsilon_{\mathcal{C}}(t)=\frac{\pi\hbar c_s(t)}{6L^2}$ is the instantaneous Casimir energy density. 
    For a fast enough drive with $\tau \leq 10$ms, an anomalous modification of the vacuum occurs. While 
    this does not manifest itself significantly on the evolution of the energy density a), the evolution of the radiative pressure is drastically altered and becomes non-monotonous b).  
    This anomalous regime occurs within the blue region of pannel c) where $-\tau^3\partial_tp$ is plotted for various values of $\tau$. 
    }
    \label{fig:6_scattering}
\end{figure}

While Eq.~\eqref{eq:MET_scattering_bis} gives us access to the anomalous correction for any velocity 
ramp profile, in this section, for simplicity, we will consider a ramp of the sound velocity of the form
\begin{equation}
    c_s(t)=c_0\left[1+\epsilon\tanh\left(\frac{t}{\tau}\right)\right]\,.
\end{equation}
The results are shown in Fig.~\ref{fig:6_scattering}. We expect the gravitational anomalies to alter the classical properties of the steady states in regions where the new energy scales $\varepsilon_\mathcal{R}$ and $\varepsilon_{\Bar{\mathcal{R}}}$ become sizable, or in other words when the sound velocity varies rapidly. Therefore, we focus in the following on the effects of the gravitational anomalies in the vicinity of the ramp at $t=0$.\\

The parameters of Fig.~\ref{fig:6_scattering} are the same that those used for Fig.~2 of the main text, motivated by the realization of a Bose-Einstein condensate in 
$^{23}$Na 
by S.~Eckel \textit{et al.} described in~\cite{eckel2018rapidly}. For a slow ramp of the condensate scattering lengths, corresponding to a ramp over a time $\tau=15ms$, $\varepsilon_{\Bar{\mathcal{R}}}$ is negligible, such that there is a single anomalous energy scale. As a consequence, the corrected energy density is approximately equal to its classical value, the instantaneous Casimir energy $\varepsilon_{\mathcal{C}}(t)$, while the pressure slightly deviates from this classical value, with a deviation quantified by $\varepsilon_\mathcal{R}$.\\
When the ramp is accelerated or in other words when $\tau$ decreases, the magnitude of both $\varepsilon_{\Bar{\mathcal{R}}}$ and $\varepsilon_{\mathcal{R}}$ increases and both the energy density and the pressure start to differ from the instantaneous Casimir energy density. While the correction to the energy density stays reasonably small as observed in Fig.~\ref{fig:6_scattering}.a, a striking effect of these anomalous corrections arises for pressure. Indeed, one observes in Fig.~\ref{fig:6_scattering} that for $\tau\lesssim10ms$, the pressure becomes non-monotonous over time, revealing once again the appearance of an instability in the system associated to the driving. On Fig.~\ref{fig:6_scattering}, the evolution of $-\tau^3\partial_tp$ reveals in greater details the emergence of this instability for $\tau\lesssim10ms$. \\

\putbib
\end{bibunit}
\end{document}